\documentclass[english, twocolumn, 10pt, aps, superscriptaddress, prb, citeautoscript]{revtex4}
\pdfoutput=1
\usepackage[utf8]{inputenc}
\usepackage[T1]{fontenc}
\usepackage{verbatim}
\usepackage{units}
\usepackage{mathtools}
\usepackage{amsmath}
\usepackage{amssymb}
\usepackage{graphicx}
\usepackage{wasysym}
\usepackage{layouts}
\usepackage{siunitx}
\usepackage{bm}
\usepackage{xcolor}
\usepackage[colorlinks, citecolor={blue!50!black}, urlcolor={blue!50!black}, linkcolor={red!50!black}]{hyperref}
\usepackage{bookmark}
\usepackage{tabularx}
\usepackage{microtype}
\usepackage[version=4]{mhchem}
\usepackage{babel}
\usepackage{float}
\usepackage{paralist}
\usepackage[normalem]{ulem}

\DeclareMathOperator{\sgn}{sgn}

\newcommand{\co}[2]{#2}

\DeclarePairedDelimiter\abs{\lvert}{\rvert}
\DeclarePairedDelimiter\norm{\lVert}{\rVert}

\makeatletter
\let\oldabs\abs
\def\abs{\@ifstar{\oldabs}{\oldabs*}}
\let\oldnorm\norm
\def\norm{\@ifstar{\oldnorm}{\oldnorm*}}
\makeatother

\newcolumntype{L}[1]{>{\raggedright\arraybackslash}p{#1}}
\newcolumntype{C}[1]{>{\centering\arraybackslash}p{#1}}
\newcolumntype{R}[1]{>{\raggedleft\arraybackslash}p{#1}}

\begin{document}
\title{Correlations in the elastic Landau level of spontaneously buckled graphene}
\author{A. L. R. Manesco}
\email{antoniolrm@usp.br}
\affiliation{Kavli Institute of Nanoscience, Delft University of Technology, P.O. Box 4056, 2600 GA Delft, The Netherlands}
\affiliation{Lorena Engineering School, University of São Paulo, Lorena, Brazil}
\author{J. L. Lado}
\affiliation{Department of Applied Physics, Aalto University, Espoo, Finland}
\author{E. V. S. Ribeiro}
\author{G. Weber}
\author{D. Rodrigues Jr}
\affiliation{Lorena Engineering School, University of São Paulo, Lorena, Brazil}

\begin{abstract}
Electronic correlations stemming from nearly flat bands in van der Waals materials have demonstrated to be
a powerful playground to engineer artificial quantum matter, including superconductors, correlated insulators and topological matter.
This phenomenology has been experimentally observed in a variety of twisted van der Waals materials, such as graphene and dichalcogenide multilayers.
Here we show that spontaneously buckled graphene can yield a correlated state, emerging from an elastic pseudo Landau level.
Our results build on top of recent experimental findings reporting that, when  placed on top of hBN or NbSe$_2$ substrates, wrinkled
graphene sheets relax forming a periodic, long-range buckling pattern.
The low-energy physics can be accurately described by electrons in the presence of a pseudo-axial gauge field, leading to the formation of sublattice-polarized Landau levels.
Moreover, we verify that the high density of states at the zeroth Landau level leads to the formation of a periodically
modulated ferrimagnetic groundstate,
which can be controlled by the application of external electric fields.
Our results indicate that periodically strained graphene is a versatile platform to explore emergent electronic states arising from
correlated elastic Landau levels.
\end{abstract}
\maketitle

\section{Introduction}

One of the key features of graphene's electronic structure is that low-energy electrons behave as massless Dirac fermions.\cite{wallaceBandTheoryGraphite1947, castronetoElectronicPropertiesGraphene2009} Among the successful applications of this model, we can highlight the prediction of the so-called zeroth Landau level (ZLL) formed exactly at the Fermi energy,\cite{novoselovTwodimensionalGasMassless2005, castronetoElectronicPropertiesGraphene2009, goerbigElectronicPropertiesGraphene2011,novoselovRoomTemperatureQuantumHall2007} in contrast to the well-known behaviour for systems with parabolic low-energy dispersion.\cite{klitzingNewMethodHighAccuracy1980, laughlinQuantizedHallConductivity1981, thoulessQuantizationParticleTransport1983} The high density of states resulting from the flat band dispersion leads to electronic instabilities at half-filling, \textit{e.g.} the formation of canted antiferromagnetic ordering in the quantum Hall edge modes.\cite{Young2012,Stepanov2018,ladoNoncollinearMagneticPhases2014,kharitonovEdgeExcitationsCanted2012, kharitonovPhaseDiagramEnsuremath2012,youngTunableSymmetryBreaking2014}

Interestingly, the emergence of Landau levels is not a unique consequence of orbital magnetic fields. They also appear when the system is subjected to \textit{pseudo} magnetic fields (PMF) and their corresponding pseudo-axial gauge fields, for example, due to the presence of strain,\cite{vozmedianoGaugeFieldsGraphene2010, amorimNovelEffectsStrains2016, levyStrainInducedPseudoMagnetic2010, morpurgoIntervalleyScatteringLongRange2006, mengStraininducedOnedimensionalLandau2013, guineaMidgapStatesCharge2008, RAMEZANIMASIR201376}
modulated interlayer hopping,\cite{PhysRevLett.108.216802} or interlayer bias.\cite{ramiresElectricallyTunableGauge2018}
In these artificial Landau levels, which also appear in twisted bilayer graphene systems displaying  the so-called magic angle flat bands bands,\cite{PhysRevB.82.121407,Bistritzer2011,PhysRevLett.108.216802,PhysRevB.99.155415} electronic instabilities are also presented.\cite{lopez-sanchoMagneticPhasesPeriodically2016, caoCorrelatedInsulatorBehaviour2018, PhysRevLett.123.096802, gonzalez-arragaElectricallyControllableMagnetism2017, viana-gomesMagnetismStrainedGraphene2009, sharmaEffectUniaxialStrain2013, codecidoCorrelatedInsulatingSuperconducting2019, caoSuperlatticeInducedInsulatingStates2016}
The emergence of correlations in van der Waals superlattices has also been reported in a variety of Moiré graphene multilayers \cite{2019arXiv191203375T,2019arXiv190308130L} and Moiré dichalcogenide multilayers,\cite{2019arXiv190703966A,2019arXiv191009047R}
suggesting that van der Waals systems combining both graphene and dichalcogenides can provide an additional new platform for correlated physics.

Here we put forward a minimal graphene-based van der Waals multilayer showing a correlated state, stemming from the emergence of localized modes associated to an elastic gauge field. Our results build on top of recent experimental reports regarding the formation of buckled graphene superlattices when the material is placed on top of hBN or NbSe$_2$  substrates.\cite{jiangFlatBandsBuckled2019, mao2020evidence} Indeed, the experimental data shows the formation of Landau subbands with sublattice polarization -- a distinctive signature of PMF, suggesting a low energy description realizing a
periodically-modulated pseudo-axial gauge field.\cite{jiangFlatBandsBuckled2019, milovanovicPeriodicallyStrainedGraphene2019, mao2020evidence} Furthermore, electronic correlations were observed when tunning the system to half-filling,\cite{mao2020evidence} consitent with the results showed in this paper.

We investigate the effects of electronic interactions in the pseudo Landau levels of buckled graphene, showing the emergence of localized correlated
states. In particular, we show the emergence of a periodically-modulated ferrimagnetic groundstate, realizing a magnetic honeycomb superlattice. We also consider the effects of charge doping, motivated by the possibility of gate-tuning graphene/hBN heterostructures, and spin-orbit coupling (SOC) due to the lack of inversion symmetry. The first shows optimal magnetization for half-filling, consistent with the correlation gap opening experimentally oberved,\cite{mao2020evidence} while the second leads to a non-collinear ferrimagnetic ordering. Finally, we show that the presence of an external perpendicular electric field breaks the global sublattice symmetry as an effective mass in the superlattice scale, suppressing the magnetic ordering when this effective mass is comparable with the gap size.

The manuscript is organized as follows. Sec. \refeq{sec:system} is devoted to introducing the studied model. In Sec. \ref{sec:magnetism}, we present key results regarding the magnetic ordering and the effects of gate-tuning, SOC and an external electric field on the interacting system. Finally, in Sec. \ref{sec:conclusion}, we summarize our results.

\section{The system}
\label{sec:system}

\begin{figure}[t!]
 \includegraphics[width=\linewidth]{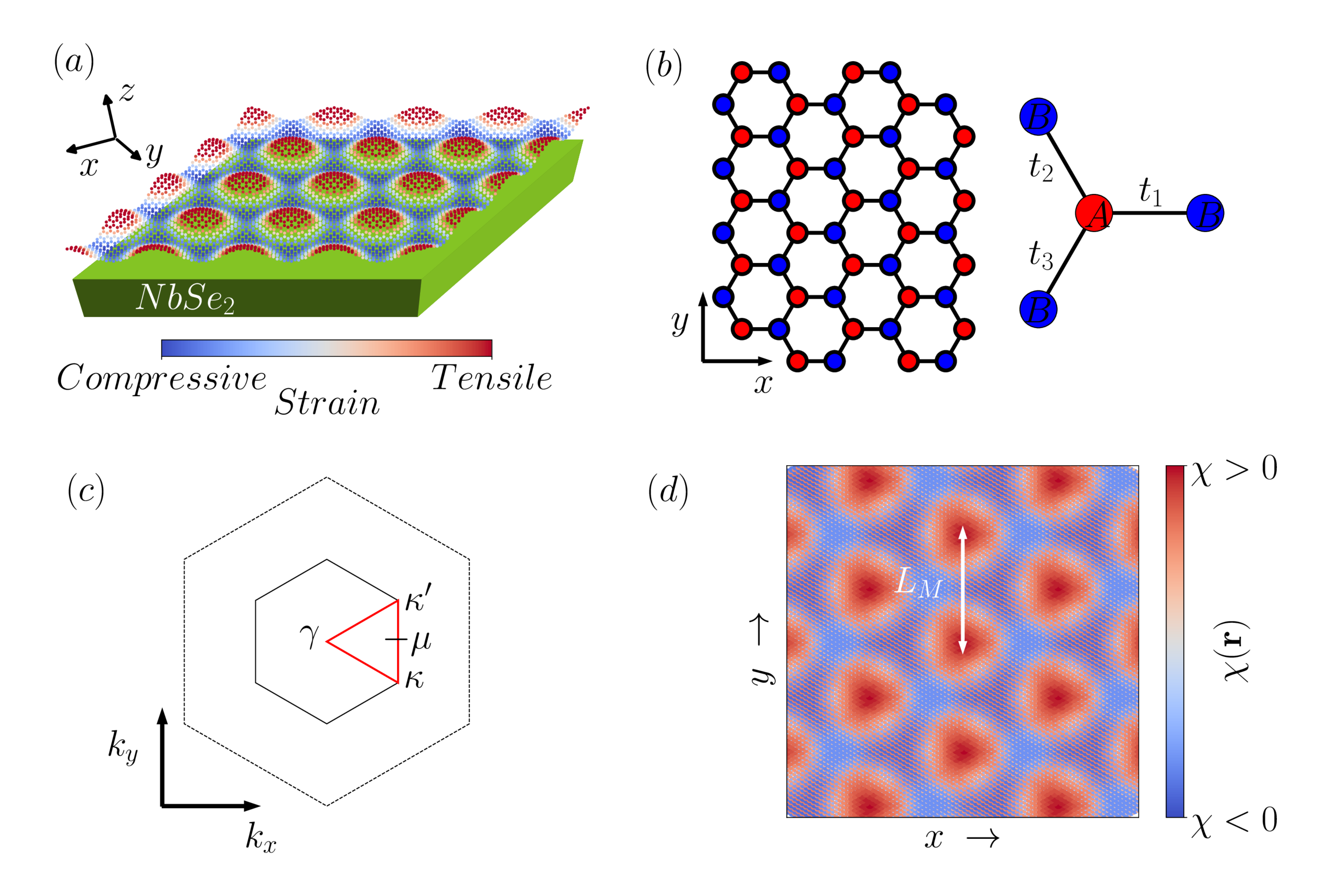}
  \caption{(a) Representation of the buckled graphene superlattice. The colors represent the magnitude of the PMF, Eq. \ref{B-field} (red is for positive and blue for negative). (b) In the presence of strain, the three hopping energies of an arbitrary atom have their degeneracy broken and we distinguish them by indexing them as $t_i$, $i \in \{1, 2, 3\}$. (c) Representation (out-of-scale) of the reduced Brillouin zone for the supercell considered in this work (small hexagon) in comparison with the larger Brillouin zone obtained with graphene's crystal vectors (large hexagon). This reduced Brillouin zone is a consequence of the long-wavelength PMF. The high-symmetry points are represented with lowercase letters to distinguish them  from graphene's original Brilluin zone high-symmetry points. (d) Valley flux for a system with a $25 \times 25$ supercell.}\label{fig:fig1}
\end{figure}

In this section, we consider a model for interacting electrons in graphene with a buckled superlattice, as depicted in Fig. \ref{fig:fig1} (a). The source code and data used to produce the figures in this work are available. \cite{antonio_l_r_manesco_2020_3703337} Our starting point is the nearest neighbour tight-binding model for a pristine graphene sheet:
\begin{equation}
\label{tb-ham}
\mathcal{H} = -t \sum_{s} \sum_{\langle i, j \rangle} c_{i s}^{\dagger}c_{j s} + U \sum_i c_{i \uparrow}^{\dagger}c_{i \uparrow}c_{i \downarrow}^{\dagger}c_{i \downarrow} ,
\end{equation}where $c^{(\dagger)}_{i s}$ annihilates (creates) electrons in the position $i$ with spin $s$. For the sake of lower computational costs, we rescaled the system according to the procedure described in \cite{liuScalableTightBindingModel2015, gonzalez-arragaElectricallyControllableMagnetism2017,PhysRevLett.123.096802}. Namely, we guarantee that the linear dispersion is preserved by fixing the Fermi velocity as $v_F \propto t a$, where $t$ is the hopping energy and $a$ is the lattice constant. Hence, we can change the lattice constant as $a \rightarrow \beta a $, as long as we compensate the hopping energy by taking $t \rightarrow t / \beta$. We consider the effects of interactions by setting a finite value for the Hubbard constant $U$,  which we rescale as $U \rightarrow U  / \beta$ to fix the ratio $U / t$. Unless explicitly written, all the results presented in this paper were obtained for a supercell with $25 \times 25$ unit cells, corresponding to $\beta \sim 2 - 3$, considering the supercell lattice constant $L_M \sim 14-18nm$.\cite{jiangFlatBandsBuckled2019} We also keep $U = t$ in all calculations, a conservative value compared to DFT estimations of $U \approx 2t $,\cite{PhysRevLett.111.036601} to ensure our results hold even if the strain is smaller than the observed values. The reduced Brillouin zone for the supercell is depicted in Fig. \ref{fig:fig1}c. The interacting Hubbard term is solved at the mean field level with the non-collinear mean field formalism.

In the presence of strain, the lattice translational symmetry is broken. Therefore, we must relax the condition that the hopping energy $t$ (and $v_F$ as well) is a constant and consider three non-equivalent hopping energies, $t_i$, $i \in \{1, 2, 3\}$, as shown in Fig. \ref{fig:fig1} (b). We can parametrize them as
\begin{equation}
\label{hopping-energies}
t_i = t + \delta t_i,
\end{equation}
with $\delta t_i$ being the difference between the new hopping energy and the one for free-standing graphene.

The low-energy description obtained by substituting Eq. \ref{hopping-energies} into Eq. \ref{tb-ham} corresponds to a modified Dirac Hamiltonian with an additional gauge field that depends on the new hoppings as:
\begin{align}
  A_x &= \frac{\sqrt{3}}{2e v_F}(t_3 - t_2),\\
  A_y &= \frac{1}{2e v_F}(t_2 + t_3 - 2t_1).
\end{align}
\co{Model we implement: references and Hamiltonian; rescaling.}
For the system under investigation, the corresponding PMF landscape was obtained by a combination of charge distribution and Landau level spacing obtained from STM and is given by:\cite{jiangFlatBandsBuckled2019, milovanovicPeriodicallyStrainedGraphene2019, mao2020evidence}
\begin{equation}
  \label{B-field}
  B({\bf r}) = B_{\text{eff}} \sum_{i=1}^3 \cos({\bf b}_i\cdot{\bf r}),
\end{equation}
with
\begin{align}
  {\bf b}_1 &= \frac{2\pi}{L_M}\left(-\frac{1}{\sqrt{3}}, 1, 0 \right),\\
  {\bf b}_2 &= \frac{2\pi}{L_M}\left(\frac{2}{\sqrt{3}}, 0, 0 \right),\\
  {\bf b}_3 &= \frac{2\pi}{L_M}\left(-\frac{1}{\sqrt{3}}, -1, 0 \right).
\end{align}
The PMF described by Eq. \ref{B-field} can be implemented by taking:
\begin{align}
\delta t_i = - \frac{\sqrt{3} e v_F L_M}{4 \pi} \sin({\bf b}_i \cdot {\bf r}).
\end{align}
Thus, the resulting gauge choice explicitly preserves the $C_3$ symmetry in the tight-binding basis. Due to the rescaling of the system, we will measure the magnetic field in terms of the dimensionless parameter $L_M / l_B$, where $l_B = \sqrt{\hbar / e B_{\text{eff}}}$. The experiment shows $L_M / l_B \sim 6 - 8$, \cite{jiangFlatBandsBuckled2019} so we take the intermediate value of $L_M / l_B = 7$ for our simulations. In fact, Fig. \ref{fig:fig3} shows that a correlated state takes place at values slightly below $L_M / l_B = 6$, even with $U = t$. Therefore, using realistic values of $U \approx 2t $\cite{PhysRevLett.111.036601} a spontaneous symmetry breaking is expected within the experimental range.

The emergence of a pseudo-axial gauge field can also be explicitly derived from the real space tight-binding model in Eq. \ref{tb-ham} with hopping constants given by Eq. \ref{hopping-energies} without resorting to the low-energy description. For this purpose, we consider the real space valley flux $\chi(\mathbf r)$ and define the valley Chern number of the system as its integral over the unit cell:
\begin{align}
C_V = C_K - C_{K'} = \int_{u.c.} \chi(\mathbf r) d^2 \mathbf r.
\end{align}
The real space valley flux in the tight binding model is equivalent to the analytically derived valley-dependent magnetic field, and therefore will reflect the real space structure
of the emergent magnetic field explicitly in the full tight binding model across the unit cell.
The real-space valley flux can be computed within the Green's function formalism as:\cite{PhysRevB.84.205137,PhysRevLett.123.096802}

\begin{equation}
    \chi(\mathbf r) =
    \langle \mathbf r  |\int_{-\infty}^0 d \omega
\int_{\scriptscriptstyle\textrm{BZ}} \frac{d^2 \mathbf k}{(2\pi)^2} \frac{\epsilon_{\alpha \beta}}{2}
 G_V (\partial_{k_\alpha}G_V^{-1}) (\partial_{k_\beta}G_V)| \mathbf r \rangle.
\end{equation}
Here, $\epsilon_{\alpha\beta}$ denotes  the Levi-Civita tensor,
\begin{align}
G_V = [\omega - H(\mathbf k)+i0^+]^{-1}
\mathcal{P}_V
\end{align}
the Green's function associated to the Bloch Hamiltonian $H(\mathbf k)$, and $\mathcal{P}_V$ the valley
polarization operator.\cite{ramiresElectricallyTunableGauge2018,PhysRevLett.120.086603,PhysRevB.99.245118}
As shown in Fig. \ref{fig:fig1} (d), it is clearly observed that certain regions of the system show a positive valley flux, whereas others have negative flux. The negative valley flux is associated with the regions with compressive anisotropic strain, whereas the positive valley flux is associated with tensile anisotropic strain. This is the very same phenomenology expected from the artificial magnetic field obtained with a low energy Dirac expansion, reinforcing the connection between the low energy model and the exact solution of the tight-binding model.

\begin{figure}[h!]
  \includegraphics[width=0.7\linewidth]{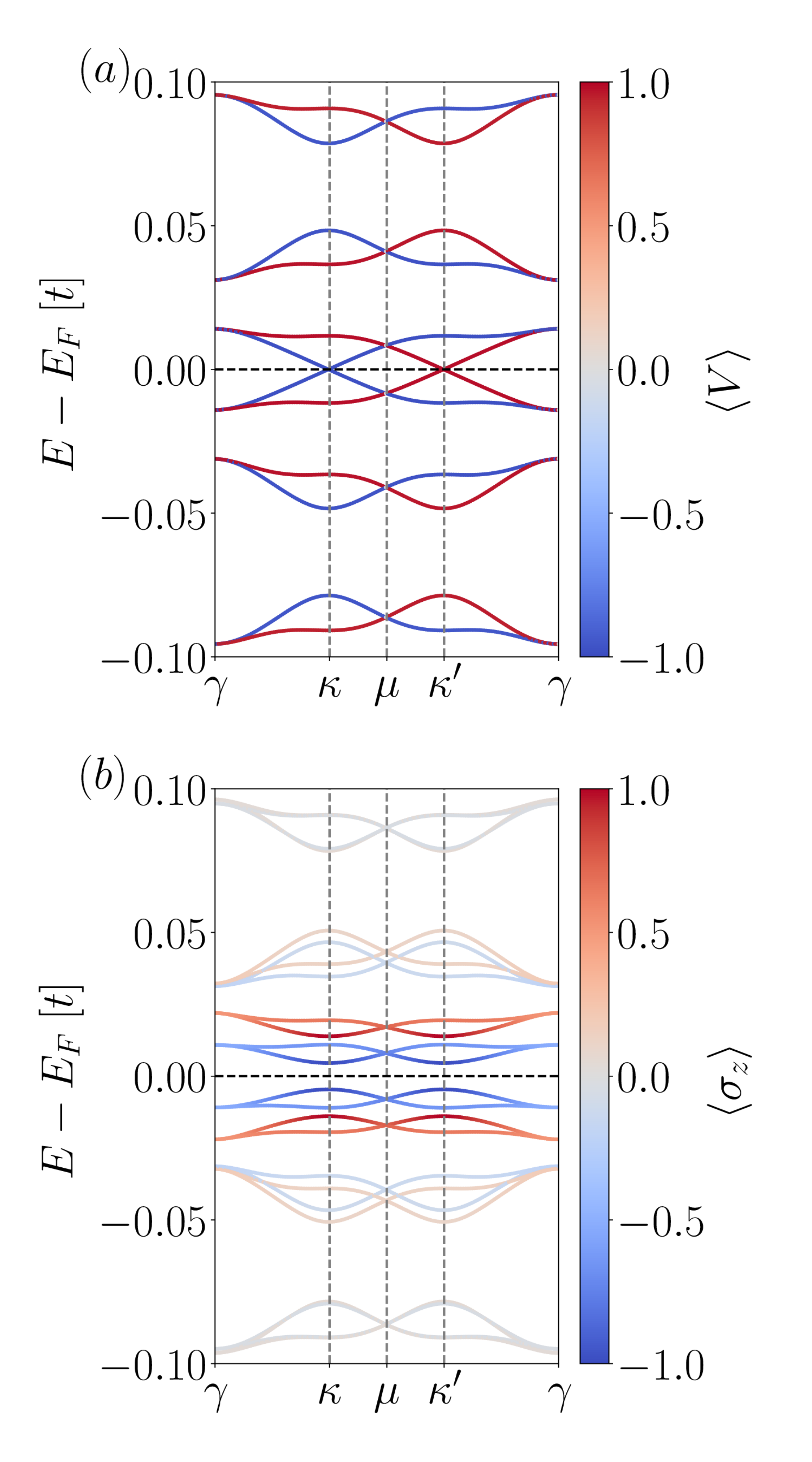}
  \caption{(a) Band diagram for a non-interacting system consisting of a periodically strained graphene sheet  projected onto the valley degrees of freedom, showing the absence of inter-valley mixing. (b) Selfconsistent band diagram for the corresponding interacting system projected onto sublattice degrees of freedom, showing an spontaneous sublattice asymmetry. The color scale in (a) indicates $+1$ for valley $K$ and $-1$ for $K'$, while in (b) $+1$ corresponds to sublattice $A$ and $-1$ to sublattice $B$.} \label{fig:fig2}
\end{figure}

\begin{figure}[t!]
  \includegraphics[width=\linewidth]{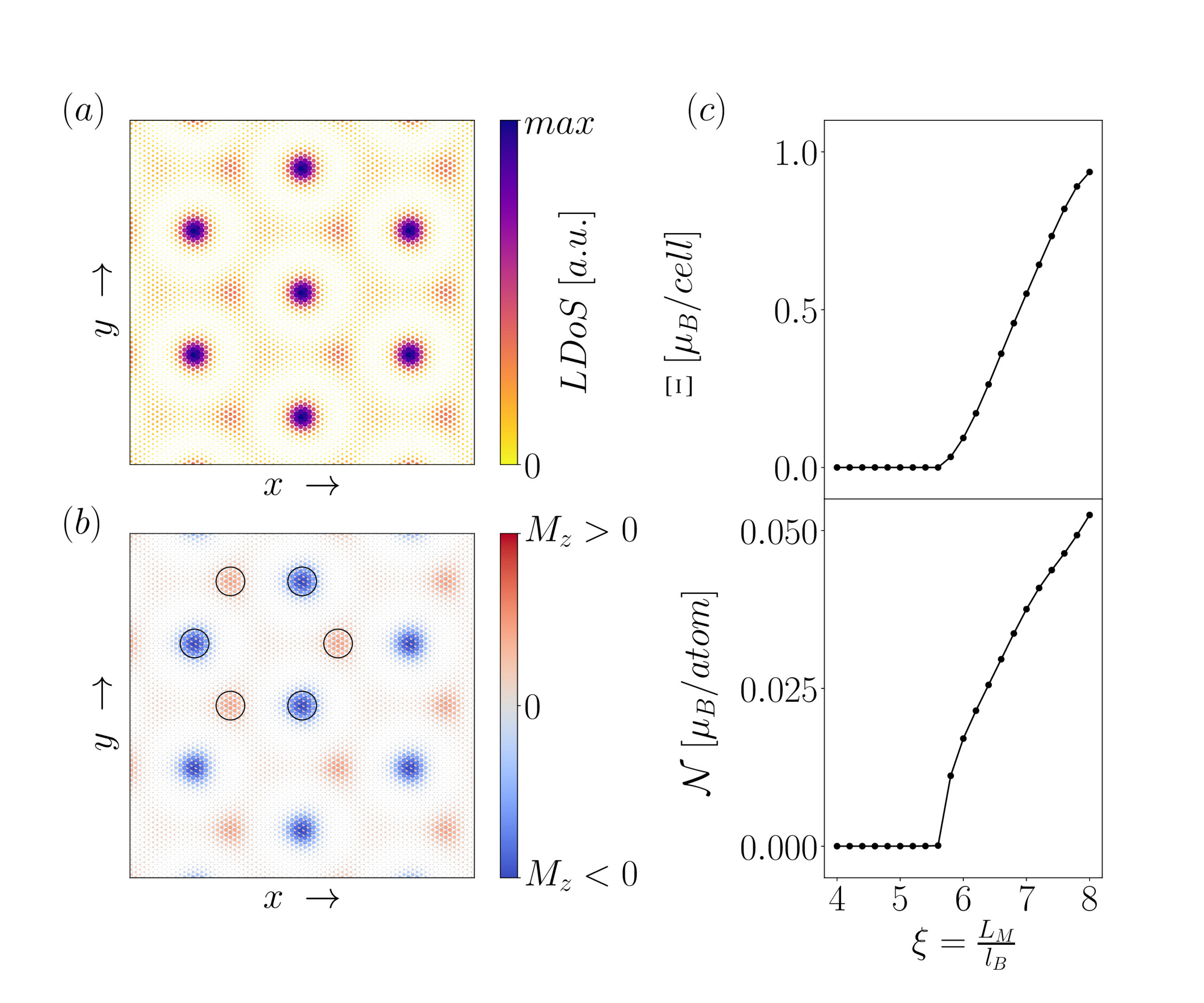}
  \caption{(a) Local density of states for the non-interacting case and (b) magnetization along the $z$-direction revealing an antiferromagnetic ordering for the interacting case. A closer analysis reveals a ferrimagnetic periodic ordering, with the ferromagnetic component changing in sign with $B(\mathbf{r})$. (c) The dependence of both the modulated ferromagnetic ($\Xi$) and antiferromagnetic ($\mathcal{N}$) order parameters on the PMF for a $10 \times 10$ supercell.} \label{fig:fig3}
\end{figure}

We now study the electronic dispersion in the absence (Fig \ref{fig:fig2} (a)) and presence (Fig \ref{fig:fig2} (b))
of electronic interactions.
In the non-interacting case, the system remains gapless even in the presence of modulated strain, but with
a highly reduced Fermi velocity due to a bandwidth quench to $W \sim 0.01t$ (Fig \ref{fig:fig2} (a)).
Moreover, we observe that the strain modulation does not create intervalley scattering
by projecting the resulting band diagram onto the valley states by means of the valley polarization operator
$\mathcal{P}_V$,\cite{ramiresElectricallyTunableGauge2018,PhysRevLett.120.086603,PhysRevB.99.245118} see Fig \ref{fig:fig2} (a). Hence, valley is still a good quantum number.
Interestingly, when interactions are turned on (Fig \ref{fig:fig2} (b)), a gap opens up in the electronic structure stemming from an emergent magnetic state that slightly breaks sublattice symmetry of the electronic spectrum. This was recently experimentally observed.\cite{mao2020evidence} We highlight that such magnetism is not expected for pristine graphene with $U=t$, and, therefore, the bandwidth quench caused by strain is essential for a correlated state to emerge, since $W\ll U$. In the following, we address in details the origin of this symmetry breaking.

\section{Magnetic ordering}
\label{sec:magnetism}

\subsection{Formation of periodically-modulated magnetization}

To better understand the emergence of the correlated state, it is convenient to look at the spatial distribution of the low energy states in the absence of interactions (Fig. \ref{fig:fig3} (a)).
The spatial distribution of these states corresponds to the zones of the superlattice under the influence of a strong elastic gauge field. Hence, according to the previous low energy discussion, these regions would be associated to zero pseudo Landau levels. A closer look also reveals that each extremum of the local density of states (LDOS) is strongly sublattice polarized, with the occupied sublattice depending on the PMF direction as expected\cite{doi:10.1021/acs.nanolett.6b04870} and experimentally observed.\cite{jiangFlatBandsBuckled2019, mao2020evidence} The localized states resulting from the buckling pattern indeed present non-zero magnetic order parameters when Hubbard interactions are considered. As expected, the magnetization (Fig. \ref{fig:fig3} (b)) correlates with the density of states of the non-interacting system (Fig. \ref{fig:fig3} (a)).

Figure \ref{fig:fig3} (b) shows the development of a periodically-modulated ferrimagnetic order parameter, which can also be interpreted as an antiferromagnetic signal with a noticeable sublattice imbalance due to the superposition of a smaller ferromagnetic signal, see Fig. \ref{fig:fig3} (c), in agreement with previous studies of a similar system.\cite{lopez-sanchoMagneticPhasesPeriodically2016} The emergence of ferrimagnetism seems counterintuitive according to Lieb's theorem for a bipartide lattice,\cite{PhysRevLett.62.1201} and, thus, requires a sublattice symmetry breaking, which in our case is provided by the PMF,\cite{doi:10.1021/acs.nanolett.6b04870} as discussed in the previous paragraph. Furthermore, even though the groundstate is ferrimagnetic, the net magnetization of the system is zero. This is also a consequence of the pseudomagnetic field, since the sublattice polarization depends on the PMF sign. In other words, it is a consequence of the sublattice symmetry being broken \emph{only} locally, but globally preserved with such PMF landscape.

In order to properly quantify the magnetization using global values within the supercell, one must carefully choose the order parameters. Since we are dealing with a system with ferrimagnetic ordering (in other words, a superposition of spatially-modulated antiferromagnetic and ferromagnetic order parameters), a good representative quantity is the standard Nèel order parameter:
\begin{equation}
  \mathcal{N} = \left| \sum_i \left( {\bf m}_i \cdot \hat z \right)  \sigma_i  \right|,
\end{equation}
where ${\bf m}_i$ is the magnetic moment at position $i$ and $ \sigma_i $ is the corresponding sublattice index $\pm 1$. It is interesting to note that the magnetization profile shown in Fig. \ref{fig:fig3} (b) corresponds to an emerging honeycomb superlattice (actually, such emerging superlattice is already visible in the non-interacting LDOS profile in Fig. \ref{fig:fig3} (a)). Namely, we can distinguish two different ferrimagnetic regions with net magnetization $M_z > 0$ and $M_z < 0$ with majority of electrons located at $A$ and $B$ sublattices, respectively. Each of these regions can be understood as different Wannier orbitals of the emerging superlattice. Therefore, defining the usual ferromagnetic order parameter $\sum_i {\bf m}_i \cdot {\bm e}_z$ is pointless, since the contribution of neighboring Wannier orbitals will cancel themselves out, leading to zero net contribution. A better idea is then to modulate the usual ferromagnetic order parameter with a function that changes in sign for different superlattice Wannier orbitals. In fact, this can be done by considering the sign of the PMF. Namely,
\begin{equation}
  \Xi = \left| \sum_i \sgn{\left[ B({\bf r}_i) \right]} \left( {\bf m}_i \cdot \hat z \right) \right|.
  \label{eq:chi}
\end{equation}
The resulting order parameter $\Xi$ is intuitively understood as the superstructure's Nèel order parameter. Thus, all the phenomenology can be reduced to the analysis of such emerging honeycomb structure. We emphasize that Eq. \ref{eq:chi} does not imply a dynamical coupling between the electronic spins and the PMF, but it reflects an indirect correlation between the magnetization and the PMF caused by sublattice polarization at the zeroth pseudo Landau level.\cite{doi:10.1021/acs.nanolett.6b04870}

We show  in Fig. \ref{fig:fig3} (c) the dependence of both magnetizations on the PMF, indicating a clear phase transition. As a matter of fact, previous results showed that the scale-independent parameter is actually the product between the PMF and the number of sites inside the supercell, \cite{lopez-sanchoMagneticPhasesPeriodically2016} a conclusion we verified for the system under consideration. The transition occurs slightly below $L_M / l_B = 6$, meaning that a magnetic groundstate is expected for PMFs within the experimental range,\cite{jiangFlatBandsBuckled2019} even using a conservative value of $U = t$, implying that for a realistic Hubbard coupling constant ($U \approx 2t$)\cite{PhysRevLett.111.036601} a spontaneous symmetry breaking is expected even for lower strain values.

\begin{figure}[t!]
  \includegraphics[width=\linewidth]{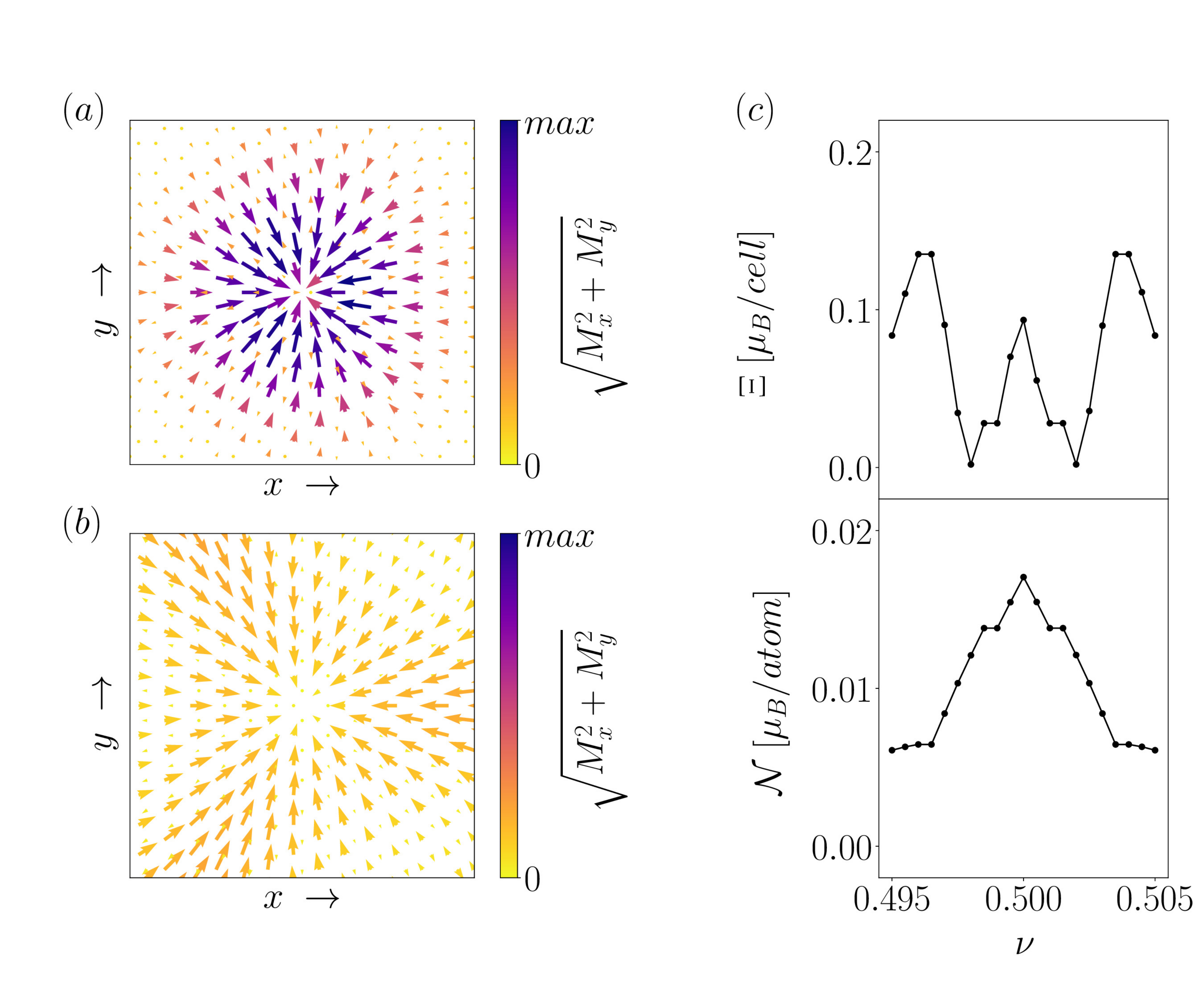}
  \caption{(a, b) In plane magnetization for both sublattices of the emerging honeycomb superlattice, showing that the groundstate presents non-collinear magnetism. In (a), we zoom in the regions corresponding to $M_z < 0$ from Fig. \ref{fig:fig3} (b), while in (b), in the regions correponding to $M_z>0$. The out-of-plane component, however, is about an order of magnitude larger and is qualitatively the same as in the case without SOC. (c) Dependency of magnetization with the filling factor for a $10 \times 10$ supercell showing that the magnetization is sensitive to doping. }\label{fig:fig4}
\end{figure}

\subsection{Effects of doping and spin-orbit coupling}

We now consider two relevant effects: \begin{inparaenum}[(i)] \item charge doping due to gate-tuning and \item effects of spin-orbit coupling (SOC) due to the break of inversion symmetry in the presence of a substrate.\end{inparaenum}

Considering charge doping, one can check in Fig. \ref{fig:fig4} (c) a rapid decay in the magnetic ordering as one goes away from half-filling. This result is consistent with the experimental observation of a gap opening as the Fermi energy approaches half-filling

We include spin-orbit coupling stemming from broken mirror symmetry with the substrate by adding a Rashba-like term to the graphene Hamiltonian:
\begin{equation}
    H_{SOC} = i \lambda_R \sum_{i,j} (\mathbf{d}_{ij} \times \mathbf{\sigma}_{s,s'} c^\dagger_{i s} c_{j s'}) \cdot \hat z,
\end{equation}
with $\lambda_R = 0.015t$ to match \textit{ab initio} estimations for graphene/NbSe$_2$ heterostructures. \cite{ganiProximityInducedSpinorbit2019}  Spin-momentum coupling explicitly breaks the spin rotation symmetry, allowing in-plane contributions to the magnetization to appear, as shown in Fig. \ref{fig:fig4} (a, b). On the other hand, the out-of plane contribution is an order of magnitude larger and qualitatively equal to calculations without SOC. Moreover, we note that in the presence of an external magnetic field,
skyrmion structures could appear in the system as a result of the combination
of the emergent Dzyaloshinskii-Moriya interaction and Zeeman field.

\subsection{Breakdown of magnetic ordering with electric fields}

\co{Effects of electric fields: gap opening and sublattice degeneracy breaking.}
The buckling pattern induces a non-homogeneous height variation of the graphene sheet with respect to the underlying substrate. Hence, the application of a perpendicular electric field should induce non-homogeneous energy shifts in real space. To account for this phenomenology, we consider the contribution from the following Hamiltonian:
\begin{equation}
  \label{eq:mu-landscape}
\mathcal{H}_{\text{elec}} = \sum_{i=1}^3 \mu({\bf r}_i) c_{i s}^{\dagger} c_{i s}\ , \quad \mu({\bf r}) = \mu_0 \sum_{i=1}^3 \cos({\bf b}_i\cdot{\bf r}),
\end{equation}
where $\mu_0$ is proportional to the amplitude of the applied electric field. It is important to emphasize a peculiar feature of the heterostructure we consider in this paper which is not expected for similar systems, \textit{e.g.} corrugated graphene.\cite{mengStraininducedOnedimensionalLandau2013}  Namely, the functional forms of $\mu({\bf r})$ (Eq. (\ref{eq:mu-landscape})) and the PMF $B({\bf r})$ (Eq. (\ref{B-field})) match. This is a direct consequence of the PMF following the topography modulation in the system.\cite{jiangFlatBandsBuckled2019} Such unique property results from a strong in-plane deformation which is hard to extract from STM measurements.\footnote{Privite communication with Slaviša Milovanović.} Indeed, this feature is essential for the electric tunnability of the system, since the nonzero values of $\mu({\bf r})$ coincide with the locations with higher local density of states.

\begin{figure}[t!]
  \includegraphics[width=\linewidth]{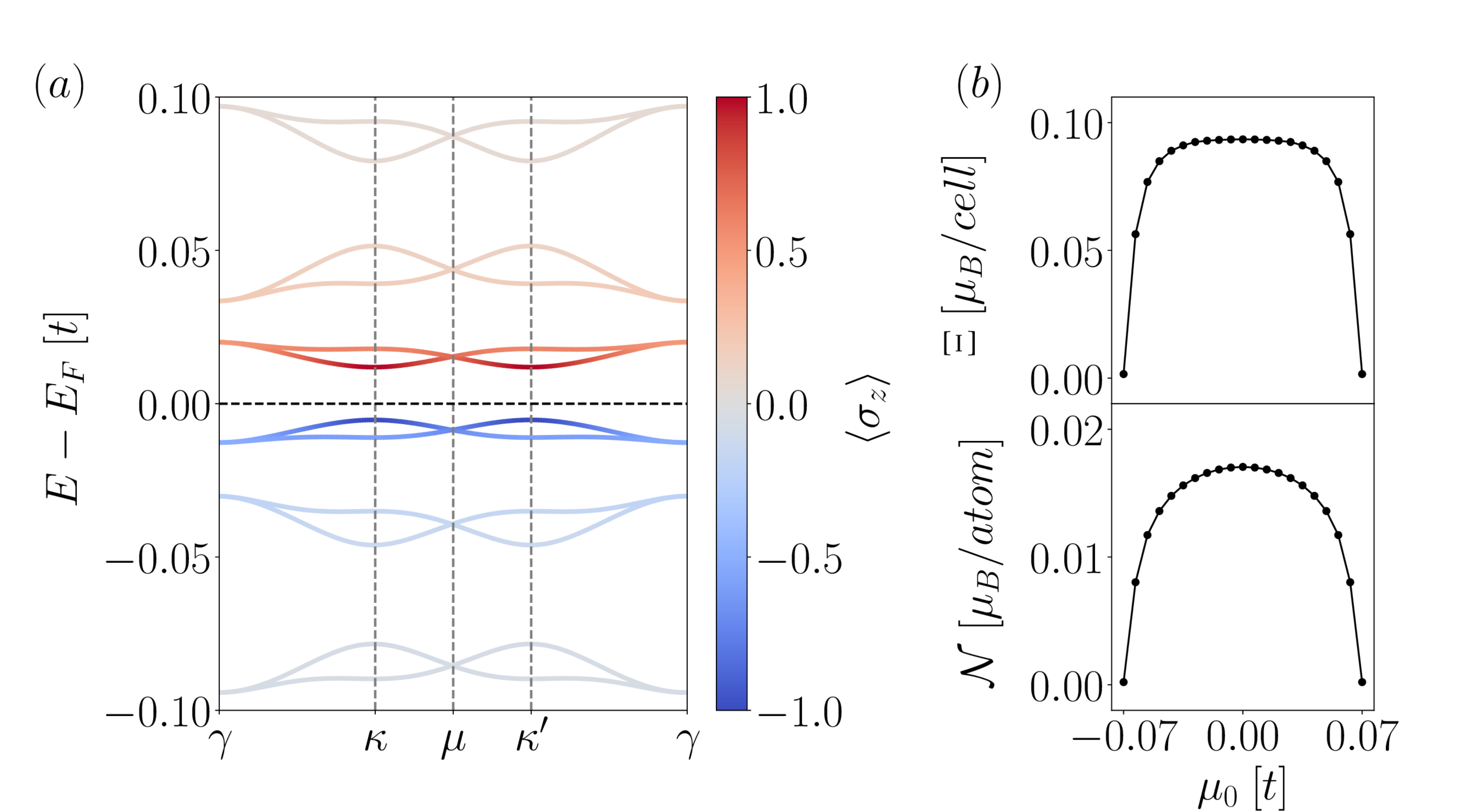}
  \caption{(a) Band structure in the presence of a perpendicular bias, showing the emergence of a gap stemming from sublattice inequivalence ($\mu_0 = 0.005t$). This bias induced gap is expected to compete with the interaction induced gap. (b) Dependence of the magnetization on the external perpendicular electric field for a $10 \times 10$ supercell.}
   \label{fig:fig5}
\end{figure}

The resulting band diagram for the noninteracting system in Fig. \ref{fig:fig5} (a) shows that, in the presence of such perpendicular electric field, a gap opens and global sublattice symmetry breaks. It is straightforward to understand this effect, if we consider the emerging honeycomb superlattice: the electric field has the same effect as a sublattice mass in the superlattice.

When electronic interactions are considered, one must expect a competition between two effects. In the presence of a finite sublattice mass, states with positive energy will be located in the sublattice $A$, while negative states, in the sublattice $B$, as one can see in Fig. \ref{fig:fig5} (a). On the other hand, in the presence of magnetic ordering, both sublattices are populated below and above the Fermi level, see Fig. \ref{fig:fig2} (b). As the sublattice mass increases, just one sublattice becomes populated below the Fermi energy, with both spin-up and spin-down states. In other words, one should expect that the magnetic ordering should be suppressed when the sublattice mass is larger than the antiferromagnetic gap. Indeed, that is exactly what we observe in Fig. \ref{fig:fig5} (b).

\section{Conclusion}
\label{sec:conclusion}

\co{We show a periodic strain-induced magnetization in graphene.}
We showed that the zeroth pseudo Landau level subband formed in buckled graphene superlattices hosts a periodic magnetically ordered groundstate at half-filling, in agreement with recent experimental results. This periodic pattern results in an emerging antiferromagnetic honeycomb superlattice. Moreover, we showed that in the non-interacting scenario, a perpendicular electric field opens up a gap and can be interpreted as a sublattice effective mass in the superlattice scale. Interestingly, in the interacting case, the competition between the bias induced mass and the antiferromagnetic gap provides a route for electrically controlling the magnetic groundstate of the system. Our results show that strained graphene provides a powerful two dimensional platform to explore correlated physics in hybrid van der Waals structures, and to study the interplay between artificial gauge fields and interactions.  Finally, it is worth noticing that the interplay of such magnetic state with the NbSe$_2$ superconductivity, not addressed in the current manuscript, can lead to a versatile platform to explore superlattice Yu-Shiba-Rusinov physics, and ultimately Majorana states.

\section*{Acknowledgements}
The authors acknowledge Slaviša Milovanović for useful discussions on the tight-binding implementation, Anton Akhmerov for useful suggestions as well as Artem Pulkin and Daniel Varjas for feedbacks on our results. The work of A.L.R.M. was funded by São Paulo Research Foundation, numbers 2016/10167-8 and 2019/07082-9. J.L.L. acknowledges the computational resources provided by the Aalto Science-IT project. A.L.R.M. also acknowledges the hospitality of the Quantum Tinkerer group.

\bibliographystyle{apsrev4-2}
\bibliography{buckled_graphene}

\end{document}